\documentclass[12pt]{spieman}  
\usepackage{amsmath,amsfonts,amssymb}
\usepackage{graphicx}
\usepackage{setspace}
\usepackage[numbers]{natbib}

\title{CASSTOR: a scientific and technology nanosatellite demonstrator for UV spectropolarimetry}

\author[a,b,*]{Coralie Neiner}
\author[a]{Vincent Lapeyrere}
\author[a]{Eitan Pechevis}
\author[a]{Claude Catala}
\author[b,a]{Boris Segret}
\author[b,c]{Feliu Lacreu}
\author[b,c]{Rashika Jain}
\author[a]{Jean-Michel Reess}
\author[d]{Fr\'ed\'eric Esteve}
\author[d]{Charles-Antoine Chevrier}
\author[d]{Jean-Luc Le Gal}
\author[d]{Laurent Doumic}
\author[c]{Adrien Saada}
\author[c]{Maelle Le Gal}
\author[a]{Olivier Dupuis}
\author[b,c]{Alexandre Dupuy}
\author[d]{Jean-Francois Aubrun}
\author[d]{Romain Pinede}
\author[d]{Aur\'elien Ledot}
\author[d]{Andr\'e Laurens}
\author[a]{Tristan Buey}

\affil[a]{LIRA, Observatoire de Paris, Universit\'e PSL, CNRS, Sorbonne Universit\'e, Universit\'e Paris Cit\'e, CY Cergy Universit\'e, 92190 Meudon, France}
\affil[b]{CENSUS, Observatoire de Paris, Universit\'e PSL,  92190 Meudon, France}
\affil[c]{LESIA, Observatoire de Paris, Universit\'e PSL, CNRS, Sorbonne Universit\'e, Universit\'e Paris Cit\'e, 92190 Meudon, France}
\affil[d]{Centre National d'\'Etudes Spatiales (CNES), Toulouse, France}

\begin{document} 
\maketitle

\begin{abstract}
In the context of the development of several space mission projects for UV spectropolarimetry at high resolution and over a wide UV wavelength range, such as Arago, Polstar, and Pollux onboard the Habitable Worlds Observatory, we are studying and developing the UV nanosatellite CASSTOR to obtain the very first UV spectropolarimetric observations of hot stars and test several new technologies, in particular a UV polarimeter and a Fine Guiding System. In this paper, we present the work and outcome of the Phase 0 study of CASSTOR.
\end{abstract}


{\noindent \footnotesize\textbf{*}Coralie Neiner,  \linkable{coralie.neiner@obspm.fr} }


\section{INTRODUCTION}
\label{sect:intro}  

\subsection{CONTEXT}

High-resolution spectropolarimetry in the visible and infrared (IR) domains has provided important insight into stellar physics and exoplanets. Going to the ultraviolet (UV) domain would greatly extend this observation technique and its scientific impact, for example for the study of circumstellar environments, star-planet interactions, grains in the interstellar medium, etc \citep[see][]{neiner2023}. Therefore, several projects for UV spectropolarimeters are currently being proposed to space agencies \citep[see][]{girardot2024}:
\begin{itemize}
    \item Pollux is a spectropolarimeter that works from the far UV at 100 nm to the near-IR at $\sim$1800 nm, with a resolution from $\sim$60000 to $\sim$120000, that is proposed by a European consortium for the Habitable Worlds Observatory (HWO) flagship mission at NASA \citep{muslimov2024}. HWO will have a 6 to 8-m diameter telescope and is expected to be launched in the early 2040s. 
    \item Polstar is a 40-cm telescope equipped with a UV spectropolarimeter that works from 113 to 278 nm with a resolution of $\sim$20000 \citep{ignace2024}. It will be proposed as a SMEX candidate at the 2026 NASA call, which would lead to a launch in 2032. 
    \item Arago is a 1-m telescope equipped with a UV and visible spectropolarimeter that works from 119 to 888 nm with a resolution of $\sim$25000 to $\sim$35000 \citep{muslimov2023}. It has been proposed as an M7 candidate to ESA but was not selected. It has recently been re-proposed at the ESA M8 call.  
\end{itemize}

\subsection{GOALS of CASSTOR}

In preparation for these future projects, we are developing a nanosatellite demonstrator called CASSTOR (CubesAt Spectropolarimeter Scientific and Technological demonstratOR). It is a 12-cm telescope equipped with a UV spectropolarimeter working from 134 to 291 nm with a resolution from $\sim$11000 to $\sim$15000 installed on a 16U platform. The goals of CASSTOR are:
\begin{enumerate}
\item demonstrate what science with high-resolution UV spectropolarimetry on a wide wavelength domain is feasible by producing the very first stellar UV spectropolarimetric measurements. To this aim, CASSTOR will observe very bright stars. In particular, it will study the magnetic field and the circumstellar environment of hot stars, including Wolf-Rayet stars, classical Be stars, and blue supergiants. No UV spectropolarimetric data of stars exist as of today, outside the Sun. CASSTOR will provide their very first stellar UV spectropolarimetric measurements.
\item help specify future larger space missions for UV spectropolarimetry, in particular Pollux for HWO. The scientific results of CASSTOR will be compared to theoretical expectations and simulations and will allow us to derive, e.g., what polarimetric precision and image stability are required for Pollux onboard HWO. The measurements of stellar calibrators with CASSTOR will also allow us to improve the demodulation process for the polarization extraction, that will subsequently be used for larger missions. 
\item demonstrate that the proposed UV polarimeter technology can withstand space conditions and provide accurate results. The CASSTOR polarimeter is made of a stack of thin birefringent plates of MgF2 in molecular adhesion followed by a MgF2 Wollaston prism. The stack of thin plates can be rotated to perform temporal modulation of the polarization, while the Wollaston prism separates the 2 orthogonally polarized beams. Such a polarimeter is the current baseline for many space mission projects for UV spectropolarimetry, such as the NUV and MUV channels of Pollux on board HWO and the Polstar SMEX proposal. Therefore, it is necessary to increase the technological readiness level (TRL) of this subsystem.
\item demonstrate that the fine guiding system of CASSTOR with an Attitude and Orbit Control System (AOCS) loop is efficient and sufficient to reach the required stability for UV spectropolarimetry and that a similar approach can be used in further missions.
\end{enumerate}

In Sect.~\ref{sect:science} we describe the science objectives of CASSTOR, its targets, and the observation strategy. In Sect.~\ref{sect:requirements} we present the requirements of CASSTOR in terms of spectroscopy, signal-to-noise ratio, and polarization errors. In Sect.~\ref{sect:design} we detail the design of the instrument proposed to reach these requirements, with particular emphasis on the new technologies, i.e. the UV polarimeter and the fine guiding system. Sect.~\ref{sect:perfos} shows the expected performances of this instrument. In Sect.~\ref{sect:budgets} we present the volume, mass, power, and data volume budgets of the payload, while in Sect.~\ref{sect:mission} we present CASSTOR's mission profile. Finally, in Sect.~\ref{sect:conclusions} we conclude on the feasibility of CASSTOR and present the expected schedule and future work for CASSTOR. 

\section{SCIENCE OBJECTIVES of CASSTOR}
\label{sect:science} 

Within galaxies, massive stars are the primary drivers of feedback and energetic processes. Throughout most of their lifetimes, stellar winds play a central role in these phenomena. However, significant uncertainties remain regarding key wind properties, such as mass-loss rates, clumping, and outflow geometry. The most reliable diagnostics of these outflows are provided by UV resonance lines, which form over large distances within stellar winds. Additionally, monitoring UV line profiles and associated linear polarisation offers insight into flow geometries. Despite this, available spectroscopic monitoring is rare and limited, and some linear polarisation studies have been hindered by insufficient sensitivity.

CASSTOR’s unique capabilities will enable a breakthrough in our understanding of stellar winds. It will refine the determination of mass-loss rates, crucial not only for constraining stellar feedback but also for modelling star–interstellar medium interactions. Furthermore, CASSTOR will help constrain the physics of winds at the critical interface between the stellar surface and the wind base, identify the origin of Discrete Absorption Components, and reveal details of wind interaction regions (e.g. effects of radiative braking, radiative inhibition, and Coriolis forces).

In approximately 10\% of hot stars, fossil magnetic fields — ranging from 100 G to several tens of kG — are observed in their radiative envelopes, typically dominated by oblique dipole configurations. These fields are believed to have formed and frozen into the star before the pre-main sequence phase. They may carry the imprint of subtle magnetohydrodynamic processes from the earliest stages of star formation, before the star becomes visible. These magnetic fields also play a significant role in subsequent evolutionary phases, particularly by influencing accretion and mass-loss mechanisms. During both main-sequence and post-main-sequence evolution, fossil fields interact strongly with stellar winds, enhancing angular momentum loss through magnetic braking. At the same time, the field impedes mass loss by redirecting outflowing material back toward the stellar surface. Since both rotation and mass loss are key to the evolution of hot stars, magnetic fields can have profound impacts.

By channelling ionised outflows, strong magnetic fields effectively control mass loss in massive stars. Notably, some of the ejected material falls back onto the star, reducing the net mass loss, but the extent of this reduction remains uncertain. Another key consequence is the braking of stellar rotation, which in turn affects stellar structure, internal mixing, chemical yields, and ultimately, stellar evolution. This phenomenon, however, is still not well understood. High-resolution UV spectropolarimetric monitoring will allow full mapping of stellar magnetospheres and provide the most sensitive diagnostics of mass loss and surface abundances. This will lead to significant advances in our understanding of transport processes in (magnetic) massive stars across all evolutionary stages, and contribute to a more complete picture of stellar feedback and the cosmic cycle of matter.

In addition, on the main sequence, classical Be stars exhibit decretion discs. A subclass of these stars — the $\gamma$ Cas objects — is suspected of amplifying dynamo-generated magnetic fields within their discs  interacting with the star. UV observations will enable unprecedented constraints on star–disc interactions in Be stars, including potential radiative ablation of the disc, and will provide much-needed insight into these enigmatic systems.

Finally, little is currently known about the magnetic fields of massive stars in the later stages of their evolution (e.g., Wolf-Rayet stars and Luminous Blue Variables). It remains difficult to observationally track the evolution of their magnetic fields over time, making it challenging to assess their influence throughout the stellar life cycle, including their possible role in the formation of enigmatic objects like magnetars. CASSTOR, with its UV spectropolarimetric capabilities, will provide critical data to address these gaps.

These scientific questions align directly with the high-priority goals of several major agencies, including the Astronomy Decadal Survey 2020, ESA’s Voyage 2050 program, Astronet’s roadmap, and CNES’s scientific prospective.

In practice, observations with CASSTOR will allow us to map in 3D the surface and environment of hot stars, e.g. their wind and disk, and draw the link between the environment and stellar surface features. We will observe bright stars at various phases of their rotation period to obtain spectropolarimetric time-series from which we can reconstruct the geometry and configuration of the surface and environment. Even though the star is observed as a point source, the spectral lines allow us to resolve the star spatially thanks to the Doppler effect. The surface can be studied through photospheric lines (as is already done in the visible domain), whereas the environment can be studied thanks to resonance lines that are only found in the UV domain. Each resonance line probes a different altitude above the surface. The maps can be produced using Zeeman-Doppler Imaging \cite[ZDI, e.g.][]{semel1989, folsom2018}. 

\subsection{TARGETS}

One of the goals of CASSTOR is to obtain the very first (outside the Sun) UV spectropolarimetric measurements of different types of stars. Only hot stars (whose radiation peaks in the UV) can be observed by CASSTOR considering its small telescope size, but these will include: magnetic hot stars, magnetic chemically peculiar stars in particular $\alpha^2$ CVn stars, OB stars, classical Be stars, Wolf-Rayet stars, blue supergiants, and magnetic supergiants. Magnetic chemically peculiar stars are considered mandatory for the mission as they are the easiest and most well-known targets and will therefore be used as prime targets for the technological demonstration. 

In addition, since we want to perform the technological demonstration of the polarimeter in all Stokes parameters (IQUV), we must observe targets that are expected to show circular polarization (Stokes V), such as magnetic stars, or linear polarization (Stokes QU), such as targets with a non-spherical environment (e.g. a disk). 

We consider that CASSTOR will have fulfilled its science goals if we can observe at least 10 stars belonging to at least 3 different types of targets, with measurements in Stokes V and QU, which would constitute a minimum sample of stars showing the various potential configurations to be tested. 

The initial target list contains 46 stars from 8 different categories: 6 $\alpha^2$ CVn stars including 2 well-known magnetic ones used as magnetic calibrators, 7 other hot magnetic stars, 7 non-magnetic, non-variable hot stars used as spectroscopic calibrators, 2 magnetic supergiants, 13 Be stars including $\gamma$ Cas, 1 Herbig Be star, 1 B[e] star, 2 Wolf-Rayet stars, and 7 blue supergiants.

\subsection{PHASE COVERAGE}

Magnetic hot stars host fossil magnetic fields that are stable in time (over decades). However, their magnetic axis is not aligned with their rotation axis. As a consequence, as the star rotates, we can see its magnetic field and magnetosphere under various inclinations. By observing the star at different rotation phases, we can then fully reconstruct the 3D map of the star and its magnetic field configuration. In the same way, over-densities in the circumstellar environment of hot stars or co-rotating interaction regions pass in the line of sight of the observer and it is possible to trace them by obtaining several observations at different rotational phases.

We thus propose that each rotationally modulated target should be observed 20 times and these observations should be spread in phase over the rotation period of the star. However, since hot stars are intrinsically stable in time, the observations can be acquired over several different rotation cycles. The rotation periods of the considered targets are known from the literature. 

From experience at visible wavelengths, we consider that if a star has been observed at least at 15 different phases, it is sufficient to produce a map, although with potential gaps or degraded spatial resolution. Conversely, more phases would improve the mapping and chances of detecting spurious events such as discrete absorption components (DACs) or Be outbursts and are thus welcome.

The CASSTOR observing plan also includes 7 non-magnetic non-variable stars, which can  be used as calibrators and for which no phase coverage is required. Multiple calibrators (these and the magnetic ones) will be observed once per month (when visible) to check and follow the drift in instrumental polarization, both in the frame of the technological demonstration and to correct the scientific data. This implies that magnetic calibrators will thus be observed first for 20 phases for magnetic characterization and then once per month for calibration. 

\section{CASSTOR REQUIREMENTS}
\label{sect:requirements}

\subsection{Spectral range}

With CASSTOR we will look at 2 different types of spectral lines: photospheric lines that provide us with information on the surface of the star, and resonance lines that are formed further out in the atmosphere and wind. Photospheric lines will be co-added with the Least-Square Deconvolution \citep[LSD, e.g.][]{donati1997, kochukhov2010} method to increase the signal-to-noise ratio (SNR). Each resonance line, however, probes a different height in the environment of the star, therefore has to be studied individually. The width of these lines is defined by the terminal velocity of the stellar wind, i.e. they are very broad ($\sim$1000 km/s) and thus cover many spectral elements. 

The width of the photospheric lines is defined mostly by the projected rotational velocity of the star ($v\sin i$), which ranges from a few km/s for magnetic chemically peculiar stars to several hundreds of km/s for classical Be stars. To obtain a sufficient SNR to measure the magnetic field at the surface of stars from photospheric lines, it is necessary to co-add as many lines as possible. The SNR of the averaged line increases at first order as the square root of the number of lines (e.g., adding 100 lines improves the SNR by a factor of $\sim$10).  

\begin{figure}
\begin{center}
\begin{tabular}{c}
\includegraphics[width=16cm]{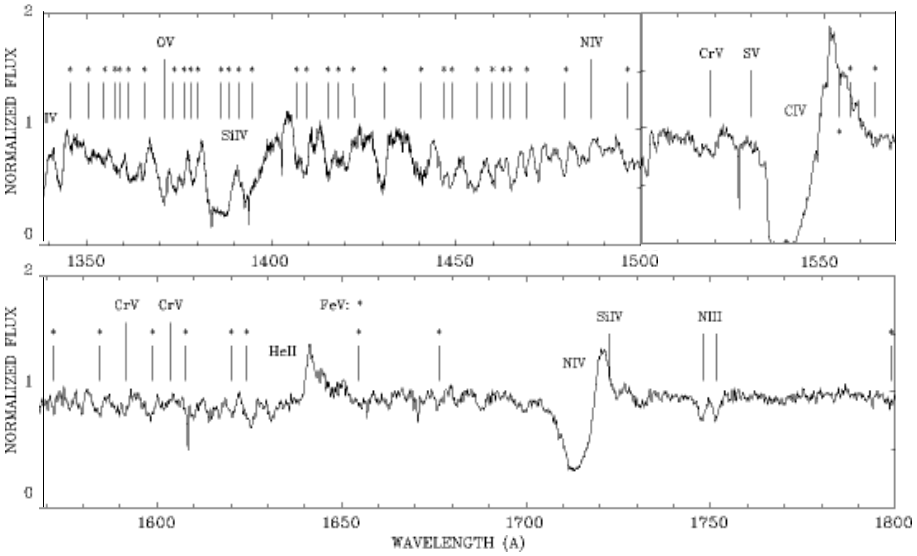}
\end{tabular}
\end{center}
\caption
{ \label{fig:zpup} Example of a merged IUE and Copernicus spectrum of the hot star $\zeta$\,Pup in the wavelength range 134-180 nm covered by CASSTOR, showing wide resonance and sharp photospheric lines. Adapted from \cite{pauldrach1994}}.
\end{figure}

The spectral range of CASSTOR has thus been defined so that it includes important resonance lines and as many photospheric lines as possible (see Fig.~\ref{fig:zpup}). We find that the optimum range is 134-291 nm. This includes the following resonance lines: SiIV doublet at 139.3 and 140.2 nm, CIV doublet at 154.8 and 155.0 nm, HeII at 164 nm, AlIII doublet at 185.4 and 186.3 nm, FeII multiplets around 234.3, 238.2 and 259.9 nm, and MgII doublet at 279.5 and 280.2 nm. In addition, the OVI line at 134.3 nm is just at the edge of the chosen wavelength domain, and a few other interesting resonance lines are just below (CII doublet at 133.4 and 133.6 nm, and OVI line at 133.8 nm), therefore it might be interesting in Phase A to try to push the wavelength range down to 132 nm (considering the width of resonance lines ($\sim$1 nm) at these wavelengths). This has not been done in the current design due to the SNR limits.

\subsection{Spectral resolution}

The spectral resolution must be high enough to sufficiently sample the profile of individual lines. Since photospheric lines are broadened primarily by rotation, the spectral resolution must correspond to a fraction of $v\sin i$ for the slowest rotators of our sample. Spectral resolution then translates into spatial resolution in the map reconstruction procedure. In order to match the $v\sin i$ of the majority of our sample stars, we have set a requirement at R=10000. 

LSD calculations performed on simulated photospheric spectra have shown that a minimum R=8000 resolution is necessary for LSD to be able to recognize and separate spectral lines in the spectrum. This resolution limit increases with increasing $v\sin i$ as the spectral lines get broader and shallower. The set requirement at R=10000 is therefore well suited for LSD. 

The resonance lines formed in the environment of the stars are usually much wider than photospheric lines and, therefore, do not require better resolution. 

\subsection{SNR}

The SNR required for magnetic field measurements in photospheric lines depends on the magnitude of the star, its $v\sin i$ i.e. the width of the lines over which the magnetic signature is spread, and its spectral type (which impacts the number and depth of the lines that can be used in the LSD calculation). For each target, it is therefore possible to define a specific SNR to reach our science goal. 

The gain between the SNR of a single line and the average SNR of all photospheric lines present in the CASSTOR wavelength range has been estimated from simulated UV spectra. We find that the gain is 130 for a Teff=10000 K star, 120 for Teff=15000 K, 100 for Teff=20000 to 30000 K, and 75 for Teff=40000 K.

From ground-based spectropolarimetry in the visible domain published in the literature, we know the polar magnetic field strength of each magnetic target. We used this value specific to each star to compute the SNR needed in the average LSD line to detect this magnetic field. From the magnitude and $v\sin i$ value of the star (available in the literature as well) and the LSD gain, we can then translate this LSD SNR value into the SNR needed in the spectrum itself. Values vary between 15 and 1126. 

For resonance lines, no UV spectropolarimetric data of hot stars have ever been obtained, and we do not know what to expect exactly in terms of the polarization signal. It is actually one goal of CASSTOR to estimate this polarization signal. However, since each line will have to be studied individually, it is necessary to have a high SNR to detect small polarization signals. Models of UV resonance lines and their polarization have been computed \citep{erba2021} and give a first estimate of a few 10$^{-3}$ intensity in the Stokes V signal for kG fields. We have thus set the required SNR to 1000. Only the first observations of CASSTOR will tell us if this is sufficient, and these results will also be very valuable for the planning of the future bigger missions of UV spectropolarimetry.

\subsection{POLARIZATION ERRORS}

The quality of a spectropolarimetric measurement does not only depend on SNR but also on the polarization error introduced by the instrument. 
To reach a satisfactory polarization measurement, the instrumental polarization must be reduced below 10$^{-3}$. As a consequence, we request the telescope to be on-axis and the polarimeter to be placed as soon as possible after the telescope and before the spectrograph. Instrumental polarization produced after the polarimeter does not influence the results (unless it significantly removes the flux of particular polarization states). 

In addition, to produce one spectropolarimetric measurement, we must obtain 4 sub-exposures taken at 4 different angles of the modulator of the polarimeter, as we must solve for 4 unknowns in the system (IQUV). The differences between the signal recorded in the 4 sub-exposures allow us to produce Stokes IQUV through a demodulation procedure \citep{pertenais2017}. Therefore, as we acquire the 4 sub-exposures, the instrument must be as stable as possible; otherwise we may produce spurious differences between the 4 sub-exposures. In particular, the position of the spectrum on the detector must not change during the series of 4 sub-exposures. The requirement on this positional stability depends on the requested maximum level of polarization error (10$^{-3}$) and on the width of the spectral line (i.e. mainly $v\sin i$). 

\begin{figure}
\begin{center}
\begin{tabular}{c}
\includegraphics[height=8cm]{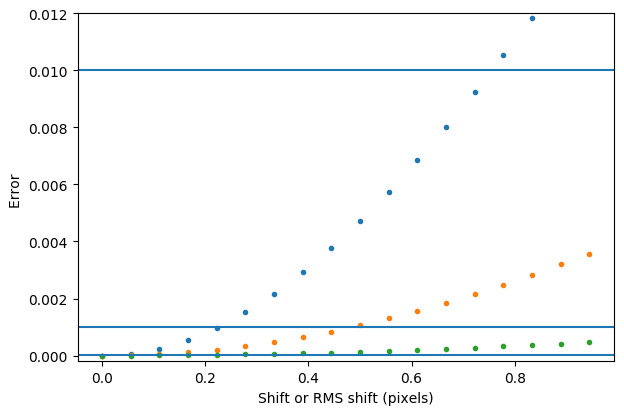}
\end{tabular}
\end{center}
\caption 
{ \label{fig:shift}
Allowed pixel shift between sub-exposures as a function of polarization error and $v\sin i$ value. The blue dotted line corresponds to $v\sin i$=15 km/s, the orange dotted line corresponds to 25 km/s, while the green dotted line corresponds to 50 km/s. The 3 horizontal solid lines indicate the 10$^{-2}$, 10$^{-3}$, and 10$^{-5}$ polarization error levels.} 
\end{figure}

Assuming a Gaussian line observed in 2 sub-exposures, we have computed the polarization error produced by a shift of the spectrum along the wavelength axis between the 2 sub-exposures (see Fig.~\ref{fig:shift}). We find that a positional stability of 2 km/s, i.e. of 0.23 pixel on the detector, is required to keep the polarization error below 10$^{-3}$ for $v\sin i$=15 km/s. An easier stability requirement is sufficient for more rapid rotators or for strong magnetic fields. Only one of the considered targets would not reach the 10$^{-3}$ polarization error with a 0.23 pixel stability. Conversely, with a less good stability, many targets could still be observed with a good polarization error due to their high $v\sin i$ or magnetic field.

\subsection{POINTING}

Finally, we will obtain several observations at different stellar rotational phases for each target, and we must be able to compare these observations to obtain a global 3D picture of the star. As a consequence, these various observations of the same star must be taken in the same polarization reference frame. In practice, this means that the Wollaston prism of the polarimeter must be oriented in the same way compared to the star, i.e. the satellite must be pointed with the same inertial attitude for all observations of a given star. 

\section{INSTRUMENT DESIGN}
\label{sect:design} 

\begin{figure}
\begin{center}
\begin{tabular}{c}
\includegraphics[height=10cm]{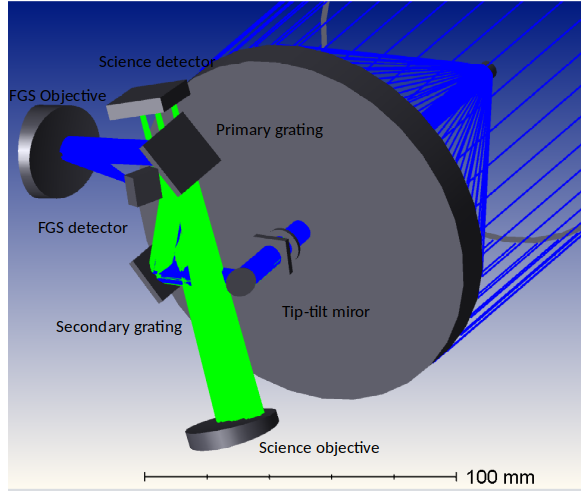}
\end{tabular}
\end{center}
\caption 
{ \label{fig:instru}
View of the instrument at the back of the primary mirror, with the science order beams. In green, we show the optical path of the science beam, going from the telescope through the modulator and Wollaston prism of the polarimeter, the tip-tilt mirror, the secondary grating, the primary grating, the science objective, and arriving on the science detector. In blue, we show the optical path of the FGS beam, following the same path as the science beam until the primary grating, but continuing afterwards to the FGS objective and FGS detector.}
\end{figure}

The CASSTOR payload is composed of a telescope, a polarimeter, a spectrograph, and a detector. In addition, it includes a Fine Guiding System (FGS) and a calibration box. A view of the instrument at the back of the primary mirror is shown in Figure~\ref{fig:instru}.

\subsection{Telescope}

\begin{figure}
\begin{center}
\begin{tabular}{c}
\includegraphics[height=6cm]{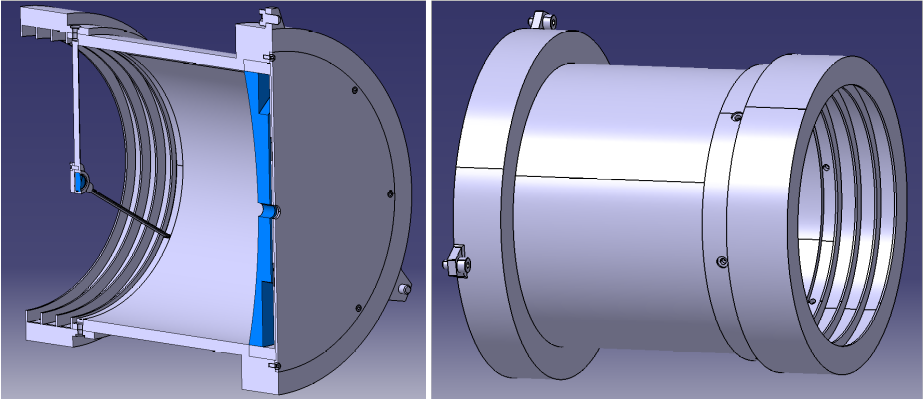}
\end{tabular}
\end{center}
\caption 
{ \label{fig:telescope}
View of the telescope model, with the mirrors in blue. Longitudinal cut (left) and full view (right).} 
\end{figure}

The telescope is an afocal telescope composed of 2 hyperbolic mirrors: a 12-cm primary mirror and a 4-mm secondary mirror. It is important to have an on-axis telescope to avoid instrumental polarization. The choice of an afocal telescope rather than a Cassegrain telescope has been made for compactness. The distance between the 2 mirrors is 10.7 cm. 
The primary mirror is open at F/1. The secondary mirror diverges. The collimated beam at the exit is 6 mm in diameter. 
The telescope is protected by a structure and a short baffle, to decrease straylight (see Fig.~\ref{fig:telescope}). The total length of the telescope assembly is 16.2 cm. At this stage, a detailed study of the straylight has not been performed and the exact size of the baffle may thus have to be adapted in the future. 

\subsection{The UV polarimeter}

The polarimeter is composed of a modulator and an analyzer. It is installed right after the telescope to avoid instrumental polarization. 

The modulator is a stack of 4 very thin (~0.3 mm) birefringent plates of MgF2 in optical contact. This stack is inserted into a rotation mechanism, which allows us to modulate the polarization of the incoming light beam with time. The thickness of each of the MgF2 plates and their respective fast axis angles have been optimized to obtain an efficiency of extraction of the polarization information of 57.7\% ($\frac{1}{\sqrt{3}}$) in all Stokes (IQUV) parameters \citep{pertenais2017} over the full wavelength range of CASSTOR. This value allows us to extract all Stokes parameters with the same efficiency. The number of angles at which the stack of plates is rotated has been set to 4, which is the minimum number of angles needed to retrieve the 4 Stokes parameters. This means that a full polarimetric measurement requires 4 sub-exposures. 

The chosen rotation mechanism is a PI RS40 V7, which can stand vacuum at 10$^{-7}$ and work between -40 and +125$^\circ$ C. Its minimal radial increment is 87 microrad, which is much better than required. This mechanism was chosen because it fits the requirements and is the most compact we could find. However, this rotation stage is not fully space-qualified. 

The analyzer is a Wollaston prism in MgF2, which separates the 2 orthogonal polarizations of the incoming beam. Therefore, there are 2 beams at the output of the analyzer that feed the spectrograph. The separation angle at the output is 0.04 degrees. 
 
The modulator is the central piece of the payload and the one that requires technological demonstration. It has been designed and tested on the ground in the frame of a R\&T (Research \& Technology) exercise funded by CNES. However, it has never flown. Although individual pieces (MgF2 plates) have flown successfully, thin MgF2 plates in optical contact have not flown. Optical contact can be fragile because MgF2 is birefringent, i.e., it has a different coefficient of thermal expansion (CTE) depending on the direction, and the MgF2 plates each have a different axis orientation. Thermal changes will thus produce different stresses in different directions, which may break the optical contact. Thermal tests have shown that $\sim$80\% of the samples pass space qualification, but some samples break when the temperature drops below -15$^\circ$ C \citep{legal2020}. As a consequence, we have set a requirement that the temperature of the polarimeter should remain within [-10:+50]$^\circ$ C.

\subsection{Spectrograph}

The echelle spectrograph is composed of a grating, a cross-disperser, and an objective mirror. The spectra cover the range 134-291 nm spread over 28 echelle orders (orders 25 to 52 of the spectrum). 
The main grating has a density of 200 lines/mm and an incidence angle of 34 degrees. The cross-disperser has a density of 600 lines/mm and an incidence angle of 35.54 degrees. The size of the gratings must be 10x8 mm and 15x15 mm, respectively, to allow us to follow the movement of the beam without vignetting until the FGS has stabilized the beam at its position. 
The objective mirror is an ellipsoid mirror of conic 0.2417 and has a focal length of 10.5 cm. Its diameter is 6 cm.

\subsection{Science detector}

We selected the Teledyne CMOS CIS-120 Capella, delta-doped by JPL \citep{nikzad2017}. It has 2048x2048 pixels of 10 microns. Its quantum efficiency (QE) varies between 0.3 and 0.55. Its readout noise is 8 $e^-$ rms. The pixel capacity is 70000 and we limited the measurements to 80\% of saturation. 

We will use this detector at a temperature of 260 K to maintain the dark level below 1.5 $e^-$/pixel/s. For a 480 s integration at 260 K, the dark level is 750 +/- 27 $e^-$. A change in temperature of 0.1 degrees results in a change in dark level of 10 $e^-$. 

One spectral element is sampled on 2.67 pixels. The final resolution of the spectrum varies between 11200 and 15000, depending on wavelength.

\subsection{Fine Guiding System}

The issue of positional stability of the spectrum in CASSTOR is addressed through the use of a Fine Guiding System (FGS) and the thermoelastic stability requirement.

The FGS relies on the recording of the 0$^{th}$ order of the spectrum on a separate small visible detector. An  objective mirror is necessary on the path of the 0$^{th}$ order beam to bring the light onto the detector. The 0$^{th}$ order means that the two gratings act as plane mirrors, i.e., there is no dispersion. Therefore, the whole light converges on a point on the FGS-detector (the flux is maximized so that it is quicker to obtain good precision on the centroid measurement). Because it is located after the polarimeter however, we record two 0-th orders, one for each polarization beam. 
 
A tip-tilt mirror, placed right behind the polarimeter, is used to maintain the 0-th order at a fixed position on the detector. This mirror has a diameter of 20 mm and can be tilted thanks to a piezo-electric system on the X and Y axes. The piezo-electric system requires a resolution of 10 $\mu$rad. Examples of such systems with a resolution down to 20 nrad and a sub-ms response time exist (e.g. from the PI (Physik Instrumente) company).

\begin{figure}
\begin{center}
\begin{tabular}{c}
\includegraphics[height=8cm]{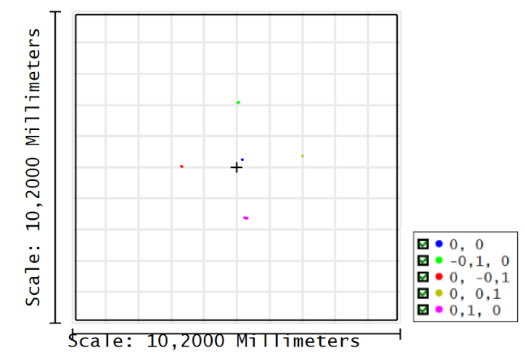}
\end{tabular}
\end{center}
\caption 
{ \label{fig:spots}
Simulation of the FGS spots on the FGS detector with off axis target by $\pm0.1^\circ$ in each direction compared to a perfectly centered spot (in blue).} 
\end{figure}

To reach the required accuracy of 0.23 arcsec (0.23 pixels) on the science detector, pixels of 1 arcsec are required on the FGS. Considering the focal length of the FGS objective, the pixels shall be 5 $\mu$m wide. The field of view of the FGS, before reaching the vignetting of the polarimeter, is about 0.1$^\circ$. To cover this field, we use a 1000 by 1000 pixels detector (see Fig.\ref{fig:spots}). As a consequence, the Absolute Pointing Error (APE) of the platform shall be below 360 arcsec off-axis.

Since the 0$^{th}$ order beam goes through the polarimeter, we record the two states of polarization of the 0$^{th}$ order, i.e. two spots on the detector separated by 73 microns ($\sim$7 pixels). Each spot is about 5x3 pixels, therefore the 2 spots partly overlap.

\begin{figure}
\begin{center}
\begin{tabular}{c}
\includegraphics[height=8cm]{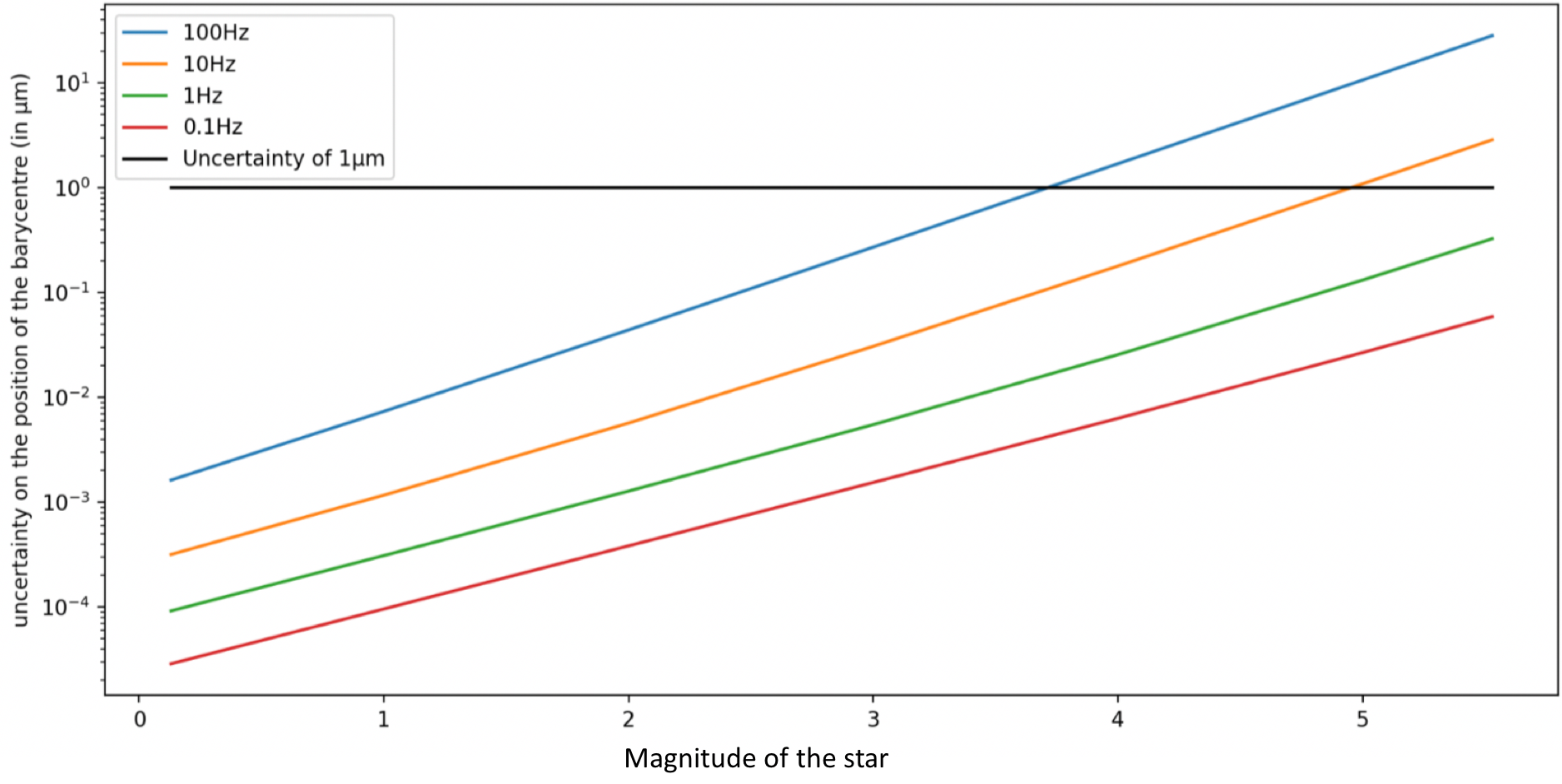}
\end{tabular}
\end{center}
\caption 
{ \label{fig:FGSperfos}
Uncertainty in the position of the spectra as function of the magnitude of the star, and the frequency of the FGS tracking loop.} 
\end{figure}

Tip-tilt in X,Y allows us to bring the spots back at the reference position on the FGS detector within $\sim$0.2 pixels (1 $\mu$m). Stabilizing the 0$^{th}$ order on the FGS detector also stabilizes the spectrum on the science detector. The requirement for spectrum stability (0.23 pixel) concerns the duration of one polarimetric measurement (currently 32 minutes maximum). The image of the 0$^{th}$ order of the spectrum has to be read with a frequency that will depend on the jitter and drift of the platform, as well as on the magnitude of the observed target. Figure~\ref{fig:FGSperfos} shows that for a magnitude V=3.5 star, the loop can run at 100 Hz, and for a V=5 star at a maximum of 10 Hz.

The tip-tilt correction introduces aberrations in the science detector images as the science beam goes through different areas of the objective. To keep these aberrations, on the edge of the spectra, lower than the required accuracy of 0.23 pixels, the tip-tilt correction shall be kept below 30 arcsec. As a consequence, the Relative Pointing Error (RPE) of the platform shall be below 30 arcsec off-axis, with feedback from the FGS measurements.

\subsection{Calibration box}
\label{sect:calibration}

The calibration box includes 2 lamps: one Deuterium lamp for flat-fielding, and one Pt-Ne hollow cathode lamp for wavelength calibration.

Deuterium lamps have their maximal intensity emission between 150 and 200 nm, which makes them suitable for CASSTOR. They have a good heritage from space missions such as IUE or HST where they were used for flat-fielding of UV spectrographs. Their typical lifetime is close to 2000 hours.

The Pt-Ne lamp has a long heritage in space missions such as IUE and HST and covers the UV spectral domain from 113 to 320 nm with more than 3000 spectral lines. It is therefore perfectly suited for CASSTOR. This hollow cathode lamp requires high voltage. According to manufacturers and the literature related to the calibration of these lamps \citep[e.g.][]{penton2008}, they require about 300 V and 10 mA for ignition and 200 V and 10 mA for steady-state operation. Depending on the platform, we may thus need to include a dedicated power supply unit for this lamp, following the precautions explained in the European cooperation for space standardization document ECSS-E-HB-20-05A\footnote{see the ECSS website at https://ecss.nl/hbstms/ecss-e-hb-20-05a-high-voltage-engineering-and-design-handbook-12-december-2012/}.

\begin{figure}
\begin{center}
\begin{tabular}{c}
\includegraphics[height=8cm]{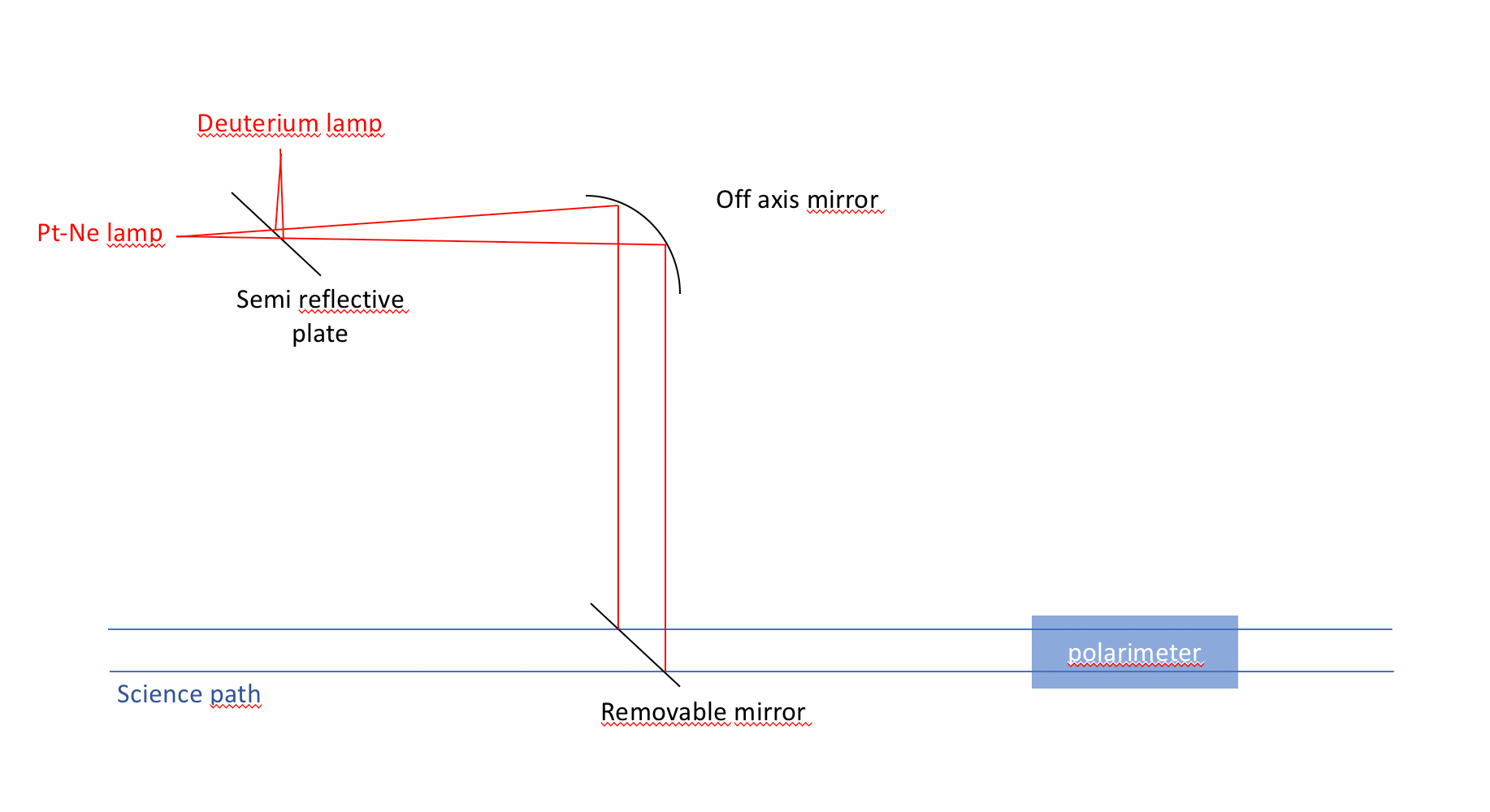}
\end{tabular}
\end{center}
\caption 
{ \label{fig:calib}
Principle of the calibration box. The two lamps are on each side of a semi-reflective plate. The science beam is shown in blue, while the beams from the 2 calibration lamps are shown in red. The removable mirror can be translated so that either the science light or the lamp light goes through the polarimeter.} 
\end{figure}

The light from one or the other lamp is selected thanks to a semi-reflective plate (see Fig.~\ref{fig:calib}). A collimating off-axis parabolic mirror forms the beam to mimic the science beam (with the same diameter). The light of the lamp is then fed to the spectropolarimeter thanks to a plane sliding mirror that enters the main beam coming from the telescope. This mirror therefore blocks the stellar light and sends the light from the lamp instead into the polarimeter. In this way the beam from the lamp follows the exact same path as the stellar light in the instrument and ends up on the same place as the science orders on the detector.

The exact design of this calibration box will depend on the total volume available on the platform. In particular, the lamps used in previous space missions were relatively big ($\sim$150 mm long). It might be necessary to make smaller versions of those for CASSTOR. A backup option is to consider a Fabry-Perot comb. 

\section{PERFORMANCES}
\label{sect:perfos} 

To evaluate the performance of the CASSTOR mission, we look at the capability of the instrument to observe the science targets. Considering the instrument described in Sect.\ref{sect:design}, a photometric budget has been calculated. 

The input of this study is the target star catalog containing the description of the 46 stars that have been selected by the science team. For each star, this catalog gives their characteristics (V magnitude, effective temperature T$_{\rm eff}$, rotation period P$_{\rm rot}$), the required SNR, and the priority of the target (1: calibrator, 2: main science target, 3: additional target).
This catalog is processed to compute the optimum exposure time for each target to do one observation. The exposure time is computed to fill 80\% of the pixel capacity. As each integration shall be repeated 4 times to perform one polarimetric measurement, we limited the maximum integration time to 8 min, so that  the sum of the 4 sub-exposures takes a maximum of 32 min and can be acquired in a single orbit. 

The second step is to compute the number of exposures $N_{\rm exp}$ required to reach the required SNR. We use:

\begin{equation}
    SNR = \frac{Sp T_{\rm exp}}{\sqrt{\sigma_r^2 + (dark + Sp) T_{\rm exp}}} \times \sqrt{N_{\rm exp}}
\end{equation}
where $Sp$ is the spectrum of the star as a function of wavelength in $e^-$/s, $\sigma_r$ is the readout noise in $e^-$, $dark$ is the dark noise in $e^-$/s, and $T_{\rm exp}$ is the exposure time in s. 

As the SNR depends on wavelength, we compute $N_{\rm exp}$ to reach the target SNR for 80\% of the spectral bins in the range [135-180] nm, where most of the photospheric spectral lines are present. The computed number of exposures is always a multiple of 4, i.e. complete polarimetric measurements. If the time spent on a single target is too long, i.e., greater than 300 h (a limit we arbitrarily set), we switch the target to "degraded science" for which we consider the SNR on the range [215-300] nm. If the required time is still above 300 h on this range, the target is discarded. 

We conclude that, out of the 46 science targets initially considered, 33 are fully accessible. For 8 targets, we can only do "degraded science". For 5 targets, the full time spent on the star for one measurement is greater than 300 hours, so these targets are discarded. We are thus left with 41 stars that can be observed by CASSTOR in its nominal mission. 

\section{BUDGETS}
\label{sect:budgets}

\subsection{Volume and Mass}

Preliminary accommodation, mass and power budgets have been performed with IDM-CIC. The study of the bus has been done at CNES PASO, based on a 16U platform.

The CASSTOR instrument fits within a volume of 170 x 170 x 243 mm$^3$. Therefore, the instrument does not fully use the width of the platform and its length is a bit longer (227 mm) than that of an 8U platform. We also recall that (1) the size of the baffle of the telescope has not been optimized and may increase once the straylight study has been performed, and (2) the current size of the calibration box is arbitrary and will depend on the size of the lamps (see Sect.~\ref{sect:calibration}). Therefore, we may need some margin in case we need a longer baffle or larger calibration box.

\begin{figure}
\begin{center}
\begin{tabular}{c}
\includegraphics[height=10cm]{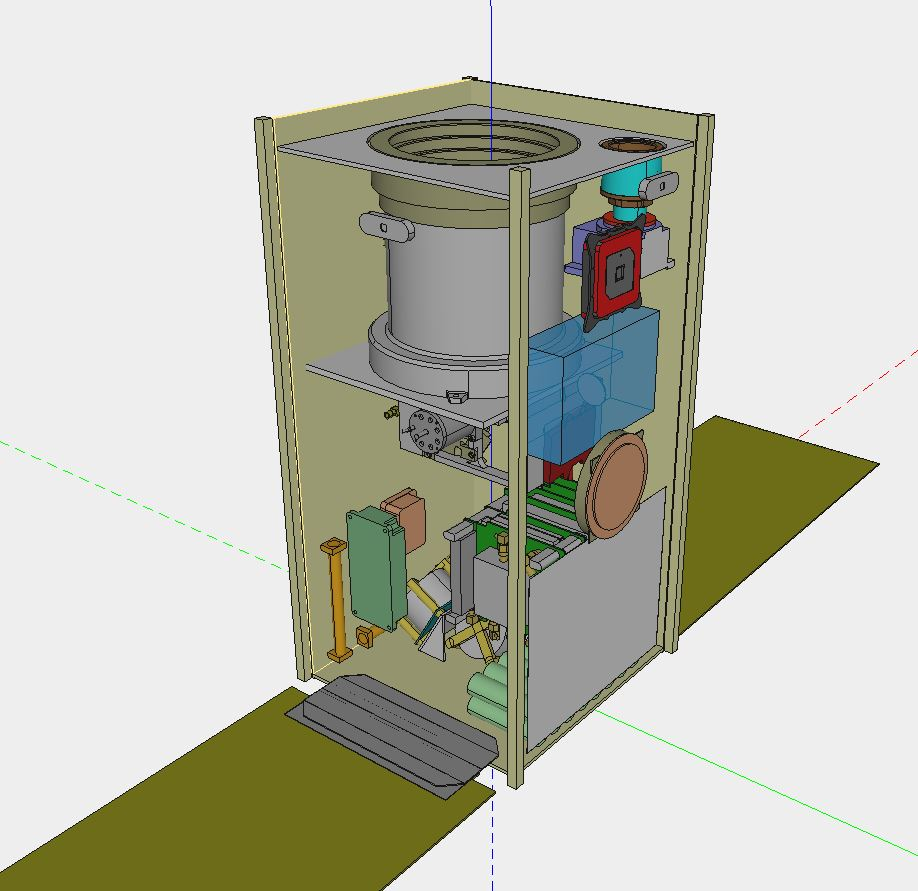}
\end{tabular}
\end{center}
\caption 
{ \label{fig:accomodation}
The CASSTOR payload accommodated with platform elements into a 16U structure. } 
\end{figure}

As shown in Fig.~\ref{fig:accomodation}, all the platform elements fit in the remaining volume of a 16U platform.

The evaluation of the mass of the payload is based on a preliminary mechanical design.
The total mass of the payload is about 6.43 kg, including margin.

\subsection{Power}

For a preliminary evaluation of the power budget, the following modes have been defined:
\begin{itemize}
\item OFF: All payload electronics are OFF.
\item STANDBY: All electronics are powered ON in standby mode. The payload data processing unit (DPU) is ON in standby mode and can receive commands. Housekeeping (HK) data are generated.
\item TRACKING: The DPU is in tracking mode. The FGS control loop is running, reading the FGS detector, and actuating the FGS tip-tilt mirror. No science acquisition. HK and FGS data are generated.
\item CALIB: The DPU is in science mode. The FGS control loop is in tracking mode, reading the FGS detector, and actuating the FGS tip-tilt mirror. The calibration source is ON. The science detector is read, and data processed by the DPU. HK, FGS, and calibration data are generated.
\item SCIENCE: the DPU is in science mode. The FGS control loop is running, reading the FGS detector, and actuating the FGS tip-tilt mirror. The calibration source is OFF. The science detector is read, and data processed by the DPU. The polarimeter rotates between each acquisition. HK, FGS and science data are generated.
\end{itemize}

Concerning the power consumption of the electronics, the following assumptions have been used:
\begin{itemize}
\item Two electronics are dedicated to control the mechanisms with the stepper motors: the rotation stage for the polarimeter, and the sliding mirror for the calibration box. These two mechanisms are not used continuously. The rotation stage rotates before each science acquisition (every few minutes). The sliding mechanism is used at the beginning and at the end of the calibration, for any new star observed. For both drivers, we considered an average power consumption of 100 mW (50 mW in standby mode).
\item The tip-tilt driver controls the piezo tip-tilt mirror. In tracking mode we considered an average consumption of 1 W (100 mW in standby mode).
\item The High Voltage calibration electronics generate the high voltage (up to 300 V) for the calibration light source. We considered an average of 2 W when the calibration box is ON (100 mW in standby mode).
\item Depending on the platform, a dedicated electronic device may be needed to produce secondary voltages. This device is not currently counted in the power budget. 
\end{itemize}

Using IDM-CIC, we produced the preliminary power budget for the different modes of the payload. The maximum power of 8.9 W is consumed in Calibration mode.

\subsection{Data volume}

The data volume of the science is evaluated without any data reduction and compression on board. The science area on the detector is 2000 by 1300 pixels. Considering 2 bytes per pixel, the size of one frame is 5 MBytes.
The full data volume of the mission can be evaluated knowing the number of frames needed for each star, which is the number of sub-exposures times the number of phases to observe. We have 28104 science frames to acquire during the full mission. This gives a total of 142.7 GBytes of data volume for a 1-year mission.
This number corresponds to an average of 24.4 MBytes per orbit.

Considering a datalink of 1 Mbit/s, the average downlink time required per orbit is 24.4$\times$8/1=195 seconds. Storage capacity will be available onboard. For example, a 10 GBytes disk storage would allow CASSTOR to wait during $\sim$25 days on average without downlink. 

Windowing and compression can be used to reduce the overall data volume and downlink time. 

\section{MISSION PROFILE}
\label{sect:mission} 

The mission needs to guarantee, on one hand, the same inertial pointing for all observations of a given star, even if these observations are separated in time, and, on the other hand, the satellite constraints in terms of power (optimization of the use of batteries), thermal aspects (necessity of a cold satellite face for as long as possible), and ADCS (desaturation of the reaction wheels) in order to maximize the availability of the satellite for science observations. To this aim, we have chosen a Sun-Synchronous Orbit (SSO) at 6/18h, a back configuration of the satellite (i.e. linked to the ground stations from the bottom face of the satellite), and an optimal attitude law. 

Using the DOCKS software suite, a model-based system engineering (MBSE) architecture has been set up for the CASSTOR project, with the aim of performing an initial mission profile analysis. We give here a description of the complete process, the inputs, and outputs. 	 

The models that are developed are \emph{not} simulators of real systems (manufactured parts or physical systems), they are intended to simulate the observation conditions and to check whether they allow scientifically useful measurements, hence to assess the system requirements.

\subsection{MBSE assumptions}

The chosen orbit for this exercise is an SSO at 6h and the altitude is fixed at 500 km. The duration of the mission has been set to 1 year, starting on January 1, 2025. Trajectories were calculated using the CNES STELA tool.

The science coverage model is adapted to an observation strategy that requires continuous observation between a star ingress and egress without any interruption (such as Earth occultation). To prevent any incident or reflected light falling on the instrument, some avoidance angles are needed to keep the line of sight always away from sources of light (Earth, Sun, Moon) present around the nanosatellite. To find the best avoidance angles, a separate in-depth study on straylight avoidance will be required in Phase A. For the Phase 0 study reported here, different values were tried based on a priori knowledge and the following values were selected: avoidance to the Sun 100$^\circ$, to the Moon 45$^\circ$, and to the Earth 25$^\circ$.

Six ground stations were examined: Kiruna (Sweden), Inuvik (Canada), New Norcia (Australia), Hartesbeesthoek (South Africa), Aussaguel (Toulouse), and Kourou (French Guiana). The intervisibility model produces the starts/ends event files for all these ground stations. Since the selected orbit is nearly polar, two ground stations are found to be the best throughout the mission: Kiruna and Inuvik. In addition, to get data during the commissioning phase, another ground station, New Norcia, is added, as it had more intervisibility than other stations during the first few weeks of the mission (South Africa was equivalent to New Norcia).

Three functional modes are considered: SCIE (science), TT\&C (telecommunication), and IDLE. The mode strategy is that priority should always be given to SCIE and then to TT\&C. TT\&C is done if the duration of the pass is more than 2 minutes and only when the on-board data volume reaches a certain limit, which is currently set to 5 GB but can be easily adjusted. In particular, this limit in data volume can be overridden if TT\&C is required for the commissioning phase. In the current model, a 5-week commissioning phase has been considered at the start of the mission, so that TT\&C will happen whenever possible in the first 5 weeks, without taking into account the on-board data volume. The model considers the selected ground stations and continues TT\&C if the stations interlace. If there is no science observation scheduled and no intervisibility with a ground station, the nanosatellite goes in IDLE mode.

In science mode, the size of one image is 2000 $\times$ 1300 pixels, and the volume of one pixel is 2 bytes, hence the volume of one image is 5078 KB (2000x1300x2/1024). The data production rate depends on the star that is being observed. For housekeeping, a data production rate of 0.2 KB/s is assumed. For TT\&C, 1 Mbit/s of data transmission rate is assumed (equivalent to an S-band). The maximum data that can be stored on-board is considered to be 10 GB.

Since no pointing is required for data download (assuming S-band communication with 2 opposite patch antennas), the functional modes are different from pointing modes. Three pointing modes are considered: SCIE (pointing to target), BARE (battery recharging mode, sun pointing), and SLEW (transition from one pointing to another). For the pointing model, +Z is considered to be the line of sight of the instrument, and the solar arrays are considered to be on the -Z plane. The nanosatellite points its solar arrays to the Sun in BARE mode and +Z to the target in SCIE mode. In SCIE mode, the quaternions are fixed according to the first observation of each target, as required by science. A slew duration of 5 minutes is assumed between two pointings and an additional tranquilization time of 5 minutes is added before each SCIE pointing, i.e., science pointing is established 5 minutes before the actual observation, so that it becomes stable before the observation.

Every 12 hours, CASSTOR stops observing science targets for 90 min ($\sim$1 orbit). This gives some time to the platform to perform dedicated activities, such as reaction wheels desaturation.

\subsection{MBSE models}

With the above assumptions, we computed MBSE models, whose results are essentially an observation plan. As inputs, the MBSE models use the catalog of targets, and the transmission and QE of CASSTOR. 
The computation of the exposure time and the number of observations required to reach the target SNR for one measurement is one of the first steps achieved by the MBSE (see Sect.~\ref{sect:perfos}).

The ingresses and egresses of all the possible observations that can be acquired for the given stars are calculated for the selected orbit, considering the ephemerids and the exclusion angles of the Sun, Moon, and Earth. 
The planning of all the observations is done by splitting the rotation period of each star into 20 phases. For each phase (which is accessible at every rotation of the star), the number of required observations shall be reached. The planning starts with the first priority targets, which are the magnetic calibrators. These targets have to be observed first 20 times to fill the 20 phases and then at least once (at a random rotational phase) every month (when possible) during the whole mission.
Once the calibrators are processed, the observations of the science stars and additional stars are planned. For each free slot in the mission schedule, an observation is planned for a star that is observable and for which the current phase is not yet fully observed. Priority criteria are defined to choose the star to be planned first if multiple targets are feasible.

The result of this planning is that among the 41 stars that are observable in terms of SNR (see Sect.~\ref{sect:perfos}), 4 stars are not observed at all. For 6 stars, the 15 phases are not fully covered. For 26 stars, more than 15 phases are fully observed, including 25 stars for which the 20 phases are fully observed. 5 stars, that are non-magnetic and non-variable, are observed at least once and, for 2 of them used as spectroscopic calibrators, a revisit is executed each month when the star is observable. Note that these results depend on the starting date of the mission, which is assumed here to be January 1, 2025.

\section{CONCLUSIONS}
\label{sect:conclusions}

The Phase 0 study of CASSTOR shows that the science coverage is good. 31 stars out of the 46 of the initial science target list (and out of the 41 that are photometrically feasible) can be observed within the requirements, and 6 additional stars can be observed with less phase coverage. The science objectives can thus be met, and CASSTOR will provide the very first UV spectropolarimetric measurements of stars outside the Sun. 

The priority calibrators can be fully observed as soon as possible at the start of the mission and then repeated every month when possible as required, which will allow for a good assessment of the instrument during flight and a technological demonstration of the UV polarimeter.

No show-stopper has been found on the technical side during this Phase 0 study. Current laboratory tests of the UV polarimeter will allow us to measure the precision that the UV polarimeter can reach. The Phase A study has recently started and focuses on testing the FGS to increase its TRL.

The development plan of CASSTOR aims at a launch at the end of 2028, to match the end of the HWO Great Observatory Maturation Program (GOMaP), in time for the selection of the Pollux instrument for HWO. 

\subsection*{Disclosures}
The authors declare that there are no financial interests, commercial affiliations, or other potential conflicts of interest that could have influenced the objectivity of this research or the writing of this paper.

\subsection* {Code, Data, and Materials Availability}

DOCKS is a software suite available at CENSUS at the Paris Observatory. It is a free and open software (https://census.psl.eu/spip.php?rubrique37\&lang=en).

IDM-CIC is a free software suite dedicated to systems engineering for space vehicle design (https://www.idm-tools.com/presentation).

STELA is an orbital propagator specialized for the Earth environment, based on a semi-analytic integration method of the orbital parameters (https://www.connectbycnes.fr/en/stela).

\subsection* {Acknowledgments}
This work has made use of the CENSUS (Center for NanoSatellites in Universe Sciences) facilities at the Paris Observatory, of the PASO service at CNES, of the SIMBAD database operated at CDS, Strasbourg (France), and of NASA's Astrophysics Data System (ADS).



\vspace{1ex}
\noindent Coralie Neiner is a CNRS director of research at the LIRA laboratory at the Paris Observatory in France. She is an expert in stellar physics, especially stellar magnetism, seismology, and hot stars. She also develops space instrumentation for UV spectropolarimetry. In particular, she is the PI of the European instrument Pollux proposed for HWO, the Arago mission proposed to ESA, and their cubesat demonstrator CASSTOR. 

Biographies and photographs of the other authors are not available.

\listoffigures

\end{document}